%% file: main.tex
\documentclass[10pt,twocolumn,letterpaper]{article}

\usepackage{iccv}
\usepackage{times}
\usepackage{epsfig}
\usepackage{graphicx}
\usepackage{amsmath}
\usepackage{amssymb}

\usepackage{subcaption}
\usepackage{multirow}
\usepackage{bm}
\usepackage{ifthen}
\usepackage[pagebackref=true,breaklinks=true,letterpaper=true,colorlinks,bookmarks=false]{hyperref}

\iccvfinalcopy 

\def\iccvPaperID{21} 

\ificcvfinal\fi

\begin{document}

\title{\vspace{-1cm}Real-ESRGAN: Training Real-World Blind Super-Resolution \\with Pure Synthetic Data\vspace{-0.2cm}}

\author{
	Xintao Wang$^{1}$ \hspace{9pt} Liangbin Xie$^{*2,3}$ \hspace{9pt} Chao Dong$^{2,4}$ \hspace{9pt} Ying Shan$^{1}$ \\
	\vspace{-0.15cm}
	\small{$^{1}$Applied Research Center (ARC), Tencent PCG} \\
	\vspace{-0.15cm}
	\small{$^{2}$Shenzhen Institutes of Advanced Technology, Chinese Academy of Sciences} \\
	\vspace{-0.15cm}
	\small{$^{3}$University of Chinese Academy of Sciences \hspace{9pt} $^{4}$Shanghai AI Laboratory} \\
	\vspace{-0.15cm}
	{\tt\small \{xintaowang, yingsshan\}@tencent.com \hspace{5pt} \{lb.xie, chao.dong\}@siat.ac.cn}\\
	\normalsize\url{https://github.com/xinntao/Real-ESRGAN}
}

\newboolean{putfigfirst}

\setboolean{putfigfirst}{true}
\ifthenelse{\boolean{putfigfirst}}{

	\twocolumn[{%
		\renewcommand\twocolumn[1][]{#1}%
		\vspace{-0.5em}
		\maketitle\thispagestyle{empty}
		\begin{center}
			\centering
			\vspace{-0.4in}
			\includegraphics[width=\linewidth]{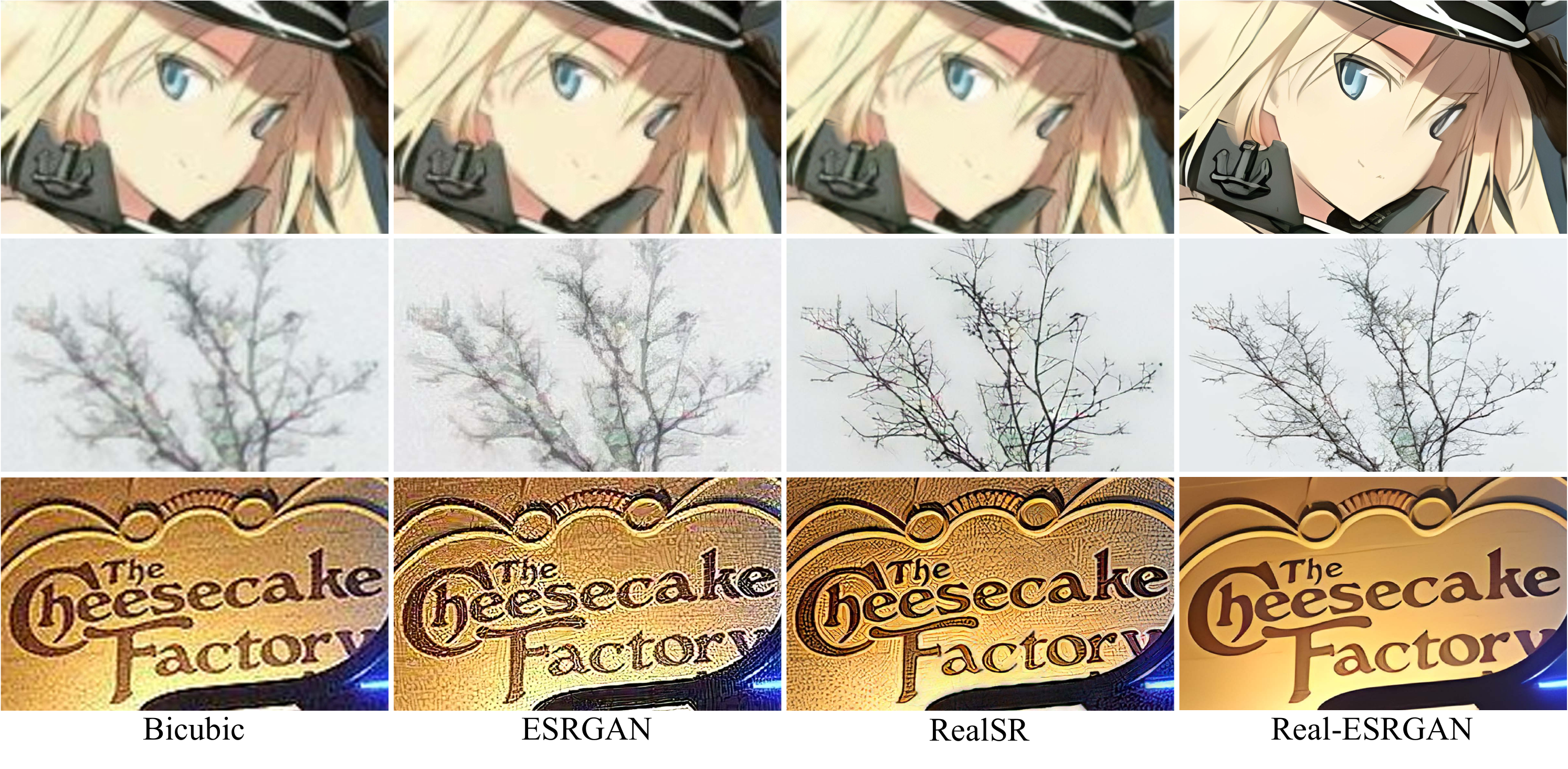}
			\vspace{-0.7cm}
			\captionof{figure}{Comparisons of bicubic-upsampled, ESRGAN~\cite{wang2018esrgan}, RealSR~\cite{ji2020real}, and our Real-ESRGAN results on real-life images. The Real-ESRGAN model trained with pure synthetic data is capable of enhancing details while removing annoying artifacts for common real-world images. (\textbf{Zoom in for best view})}
			\label{fig:teaser}
		\end{center}%
	}]
}
{
	\maketitle
}

\begin{abstract}
\let\thefootnote\relax\footnotetext{*Liangbin Xie is an intern in Applied Research Center, Tencent PCG}
\vspace{-0.4cm}Though many attempts have been made in blind super-resolution to restore low-resolution images with unknown and complex degradations, they are still far from addressing general real-world degraded images.
In this work, we extend the powerful ESRGAN to a practical restoration application (namely, Real-ESRGAN), which is trained with pure synthetic data.
Specifically,
a high-order degradation modeling process is introduced to better simulate complex real-world degradations.
We also consider the common ringing and overshoot artifacts in the synthesis process.
In addition, we employ a U-Net discriminator with spectral normalization to increase discriminator capability and stabilize the training dynamics.
Extensive comparisons have shown its superior visual performance than prior works on various real datasets.
We also provide efficient implementations to synthesize training pairs on the fly.
\end{abstract}

\input{sections/1_introduction}

\input{sections/2_related_work}
\input{sections/3_method}

\input{sections/4_experiments}

\section{Conclusion}
\vspace{-0.2cm}
In this paper, we train the practical Real-ESRGAN for real-world blind super-resolution with pure synthetic training pairs.
In order to synthesize more practical degradations, we propose a high-order degradation process and employ $sinc$ filters to model common ringing and overshoot artifacts.
We also utilize a U-Net discriminator with spectral normalization regularization to increase discriminator capability and stabilize the training dynamics.
Real-ESRGAN trained with synthetic data is able to enhance details while removing annoying artifacts for most real-world images.
\vspace{0.2cm}

\noindent\textbf{Acknowledgement.} This work is partially supported by National Natural Science Foundation of China (61906184), the Shanghai Committee of Science and Technology, China (Grant No. 21DZ1100800 and 21DZ1100100).


{\small
\bibliographystyle{ieee_fullname}
\bibliography{bib}
}

\clearpage
\appendix
\input{sections/5_appendix}

\end{document}

%% file: sections/1_introduction.tex
\section{Introduction}

Single image super-resolution (SR)~\cite{glasner2009super,dong2016image,lim2017enhanced} is an active research topic, which aims at reconstructing a high-resolution (HR) image from its low-resolution (LR) counterpart.
Since the pioneering work of SRCNN~\cite{dong2014learning}, deep convolution neural network (CNN) approaches have brought prosperous developments in the SR field. 
However, most approaches~\cite{kim2016accurate,lim2017enhanced,johnson2016perceptual,ledig2017photo,wang2018esrgan} assume an \textit{ideal bicubic downsampling kernel}, which is different from real degradations. This degradation mismatch makes those approaches unpractical in real-world scenarios. 

Blind super-resolution~\cite{michaeli2013nonparametric,bell2019blind,zhang2018learning}, on the contrary, aims to restore low-resolution images suffering from \textit{unknown and complex degradations}. 
Existing approaches can be roughly categorized into \textit{explicit modeling} and \textit{implicit modeling}, according to the underlying degradation process. 
Classical degradation model~\cite{elad1997restoration,liu2013bayesian}, which consists of blur, downsampling, noise and JPEG compression (more details in Sec.~\ref{sec:classical_degradation_model}), is widely adopted in explicit modeling methods~\cite{zhang2018learning,gu2019blind,luo2020unfolding}. 
However, the real-world degradations are usually too complex to be modeled with a simple combination of multiple degradations. Thus, these methods will easily fail in real-world samples.
Implicit modeling methods~\cite{yuan2018unsupervised,FritscheGT19,wang2021unsupervised} utilize data distribution learning with Generative Adversarial Network (GAN)~\cite{goodfellow2014generative} to obtain the degradation model. Yet, they are limited to the degradations within training datasets, and could not generalize well to out-of-distribution images. 
Readers are encouraged to refer to a recent blind SR survey~\cite{liu2021blindsurvey} for a more comprehensive taxonomy.

In this work, we aim to extend the powerful ESRGAN~\cite{wang2018esrgan} to restore general real-world LR images by synthesizing training pairs with a more practical degradation process.
The real complex degradations usually come from \textit{complicate combinations of different degradation processes}, such as imaging system of cameras, image editing, and Internet transmission.
For example, when we take a photo with our cellphones, the photos may have several degradations, such as camera blur, sensor noise, sharpening artifacts, and JPEG compression. 
We then do some editing and upload to a social media app, which introduces further compression and unpredictable noises.
The above process becomes more complicated when the image is shared several times on the Internet. 

This motivates us to extend the \textit{classical ``first-order'' degradation model} to \textbf{``high-order'' degradation modeling} for real-world degradations, \ie, the degradations are modeled with several repeated degradation processes, each process being the classical degradation model.
Empirically, we adopt a \textit{second-order degradation process} for a good balance between simplicity and effectiveness.
A recent work~\cite{zhang2021designing} also proposes a random shuffling strategy to synthesize more practical degradations. However, it still involves a fixed number of degradation processes, and whether all the shuffled degradations are useful or not is unclear.
Instead, high-order degradation modeling is more flexible and attempts to mimic the real degradation generation process.
We further incorporate $sinc$ filters in the synthesis process to simulate the \textbf{common ringing and overshoot artifacts}.

As the degradation space is much larger than ESRGAN, the training also becomes challenging. 
Specifically, 1) the discriminator requires a more powerful capability to discriminate realness from complex training outputs, while the gradient feedback from the discriminator needs to be more accurate for local detail enhancement.
Therefore, we improve the VGG-style discriminator in ESRGAN to an \textbf{U-Net design}~\cite{schonfeld2020u,yan2021fine,ronneberger2015u}.
2) The U-Net structure and complicate degradations also increase the training instability. Thus, we employ the \textbf{spectral normalization (SN) regularization}~\cite{miyato2018spectral,schonfeld2020u} to stabilize the training dynamics.
Equipped with the dedicated improvements, we are able to easily train our Real-ESRGAN and achieve a good balance of local detail enhancement and artifact suppression.

To summarize, in this work,
\textbf{1)} we propose a high-order degradation process to model practical degradations, and utilize $sinc$ filters to model common ringing and overshoot artifacts.
\textbf{2)} We employ several essential modifications (\eg, U-Net discriminator with spectral normalization) to increase discriminator capability and stabilize the training dynamics.
\textbf{3)} Real-ESRGAN trained with pure synthetic data is able to restore most real-world images and achieve better visual performance than previous works, making it more practical in real-world applications.

%% file: sections/2_related_work.tex
\section{Related Work}
\noindent\textbf{The image super-resolution} field~\cite{kim2016accurate,lai2017deep,timofte2017ntire,haris2018deep,ledig2017photo,lim2017enhanced,zhang2018residual,kim2016deeply,tai2017image,zhang2018rcan,dai2019second,liu2018non} has witnessed a variety of developments since SRCNN~\cite{dong2014learning,dong2016image}.
To achieve visually-pleasing results, generative adversarial network~\cite{goodfellow2014gan} is usually employed as loss supervisions to push the solutions closer to the natural manifold~\cite{ledig2017srgan,sajjadi2017enhancenet,wang2018esrgan,wang2018sftgan}.
Most methods assume a bicubic downsampling kernel and usually fail in real images.
Recent works also incorporate reinforcement learning or GAN prior to image restoration~\cite{yu2019path,chan2020glean,wang2021gfpgan}.

There have been several excellent explorations in blind SR. 
The first category involves explicit degradation representations and typically consists of two components: degradation prediction and conditional restoration.
The above two components are performed either separately~\cite{bell2019blind, zhang2018learning} or jointly (iteratively)~\cite{gu2019blind, luo2020unfolding, wang2021unsupervised}.
These approaches rely on predefined degradation representations (\eg, degradation types and levels), and usually consider simple synthetic degradations.
Moreover, inaccurate degradation estimations will inevitably result in artifacts.

Another category is to obtain/generate training pairs as close to real data as possible, and then train a unified network to address blind SR.
The training pairs are usually 1) captured with specific cameras followed by tedious alignments~\cite{cai2019toward,wei2020cdc}; 2) or directly learned from unpaired data with cycle consistency loss~\cite{yuan2018unsupervised,lugmayr2019unsupervised}; 3) or synthesized with estimated blur kernels and extracted noise patches~\cite{zhou2019kernel,ji2020real}.
However, 1) the captured data is only constrained to degradations associated with specific cameras, and thus could not well generalize to other real images; 2) learning fine-grained degradations with unpaired data is challenging, and the results are usually unsatisfactory.

\noindent\textbf{Degradation models.}
Classical degradation model~\cite{elad1997restoration,liu2013bayesian} is widely adopted in blind SR methods~\cite{zhang2018learning,gu2019blind,luo2020unfolding}. 
Yet, real-world degradations are usually too complex to be explicitly modeled.
Thus, implicit modeling attempts to learn a degradation generation process within networks~\cite{yuan2018unsupervised,FritscheGT19,wang2021unsupervised}.
In this work, we propose a flexible high-order degradation model to synthesize more practical degradations. 

%% file: sections/3_method.tex

\section{Methodology}

\subsection{Classical Degradation Model}
\label{sec:classical_degradation_model}
Blind SR aims to restore high-resolution images from low-resolution ones with unknown and complex degradations.
The classical degradation model~\cite{elad1997restoration,liu2013bayesian} is usually adopted to synthesize the low-resolution input.
Generally, the ground-truth image $\bm{y}$ is first convolved with blur kernel $\bm{k}$. Then, a downsampling operation with scale factor $r$ is performed. The low-resolution $\bm{x}$ is obtained by adding noise $\bm{n}$. Finally, JPEG compression is also adopted, as it is widely-used in real-world images.
\vspace{-0.2cm}
\begin{equation}\label{equ:degradation}
	\vspace{-0.2cm}
	\bm{x} = \mathcal{D}(\bm{y}) =  [(\bm{y}\circledast \bm{k})\downarrow_{r} + \bm{n}]_{\mathtt{JPEG}},
\end{equation}
where $\mathcal{D}$ denotes the degradation process. In the following, we briefly revisit these commonly-used degradations. The detailed settings are specified in Sec.~\ref{sec:implementation}.  More descriptions and examples are in Appendix.~\ref{sec:more_detail_degradation_model}.
\newline\newline
\noindent\textbf{Blur.}
We typically model blur degradation as a convolution with a linear blur filter (kernel).
Isotropic and anisotropic Gaussian filters are common choices.
For a Gaussian blur kernel $\bm{k}$ with a kernel size of $2t+1$, its $(i,j) \in [-t, t]$ element is sampled from a Gaussian distribution, formally:
\vspace{-0.2cm}
\begin{align}\label{equ:blur}
	\vspace{-0.4cm}
	\bm{k}(i,j) &= \frac{1}{N} \exp(-\frac{1}{2}\bm{C}^T\bm{\Sigma}^{-1}\bm{C}), \quad \bm{C}=[i,j]^T,
\end{align}
where $\bm{\Sigma}$ is the covariance matrix; $\bm{C}$ is the spatial coordinates; $N$ is the normalization constant.
The covariance matrix could be further represented as follows:
\vspace{-0.2cm}
\begin{align}\label{equ:cov_matrix}
	\vspace{-0.2cm}
	\bm{\Sigma} &= \bm{R} \begin{bmatrix}
		\sigma_1^2 & 0 \\
		0 & \sigma_2^2
	\end{bmatrix} \bm{R}^T, \quad \text{($\bm{R}$ is the rotation matrix)}\\
	&=\begin{bmatrix}
		cos\theta& -sin\theta \\
		sin\theta & cos\theta
	\end{bmatrix}\begin{bmatrix}
		\sigma_1^2 & 0 \\
		0 & \sigma_2^2
	\end{bmatrix} \begin{bmatrix}
	cos\theta& sin\theta \\
	-sin\theta & cos\theta
\end{bmatrix},
\end{align}
where $\sigma_1$ and $\sigma_2$ are the standard deviation along the two principal axes (\ie, eigenvalues of the covariance matrix); $\theta$ is the rotation degree.
When $\sigma_1 = \sigma_2$, $\bm{k}$ is an isotropic Gaussian blur kernel; otherwise  $\bm{k}$ is an anisotropic kernel.

\noindent\textbf{Discussion.}
Though Gaussian blur kernels are widely used to model blur degradation,
they may not well approximate real camera blur.
To include more diverse kernel shapes, we further adopt generalized Gaussian blur kernels~\cite{liu2020estimating} and a plateau-shaped distribution. Their probability density function (pdf) are $ \frac{1}{N} \exp(-\frac{1}{2} (\bm{C}^T\bm{\Sigma}^{-1}\bm{C})^\beta$, and $\frac{1}{N} \frac{1}{1+(\bm{C}^T\bm{\Sigma}^{-1}\bm{C})^\beta}$, respectively. $\beta$ is the shape parameter.
Empirically, we find that including these blur kernels could produce sharper outputs for several real samples.
\newline\newline
\noindent\textbf{Noise.}
We consider two commonly-used noise types -- 1) additive Gaussian noise and 2) Poisson noise.
Addictive Gaussian noise has a probability density function equal to that of the Gaussian distribution.
The noise intensity is controlled by the standard deviation (\ie, sigma value) of the Gaussian distribution.
When each channel of RGB images has independent sampled noise, the synthetic noise is color noise.
We also synthesize gray noise by employing the same sampled noise to all three channels~\cite{zhang2021designing,nam2016holistic}.

Poisson noise follows the Poisson distribution. It is usually used to approximately model the sensor noise caused by statistical quantum fluctuations, that is, variation in the number of photons sensed at a given exposure level.
Poisson noise has an intensity proportional to the image intensity, and the noises at different pixels are independent.
\newline\newline
\noindent\textbf{Resize (Downsampling).}
Downsampling is a basic operation for synthesizing low-resolution images in SR.
More generally, we consider both downsamping and upsampling, \ie, the resize operation.
There are several resize algorithms -  nearest-neighbor interpolation, area resize, bilinear interpolation, and bicubic interpolation. Different resize operations bring in different effects - some produce blurry results while some may output over-sharp images with overshoot artifacts.

In order to include more diverse and complex resize effects, we consider a random resize operation from the above choices. As nearest-neighbor interpolation introduces the misalignment issue, we exclude it and only consider the area, bilinear and bicubic operations.
\newline\newline
\noindent\textbf{JPEG compression.}
JPEG compression is a commonly used technique of lossy compression for digital images.
It first converts images into the YCbCr color space and downsamples the chroma channels.
Images are then split into $8\times 8$ blocks and each block is transformed with a two-dimensional discrete cosine transform (DCT), followed by a quantization of DCT coefficients. More details of JPEG compression algorithms can be found in~\cite{shin2017jpeg}.
Unpleasing block artifacts are usually introduced by the JPEG compression.

The quality of compressed images is determined by a quality factor $q\in [0, 100]$, where a lower $q$ indicates a higher compression ratio and worse quality.
We use the PyTorch implementation - $\mathtt{DiffJPEG}$~\cite{diffjpeg}.

\begin{figure*}
	\vspace{-0.8cm}
	\begin{center}
		\includegraphics[width=\linewidth]{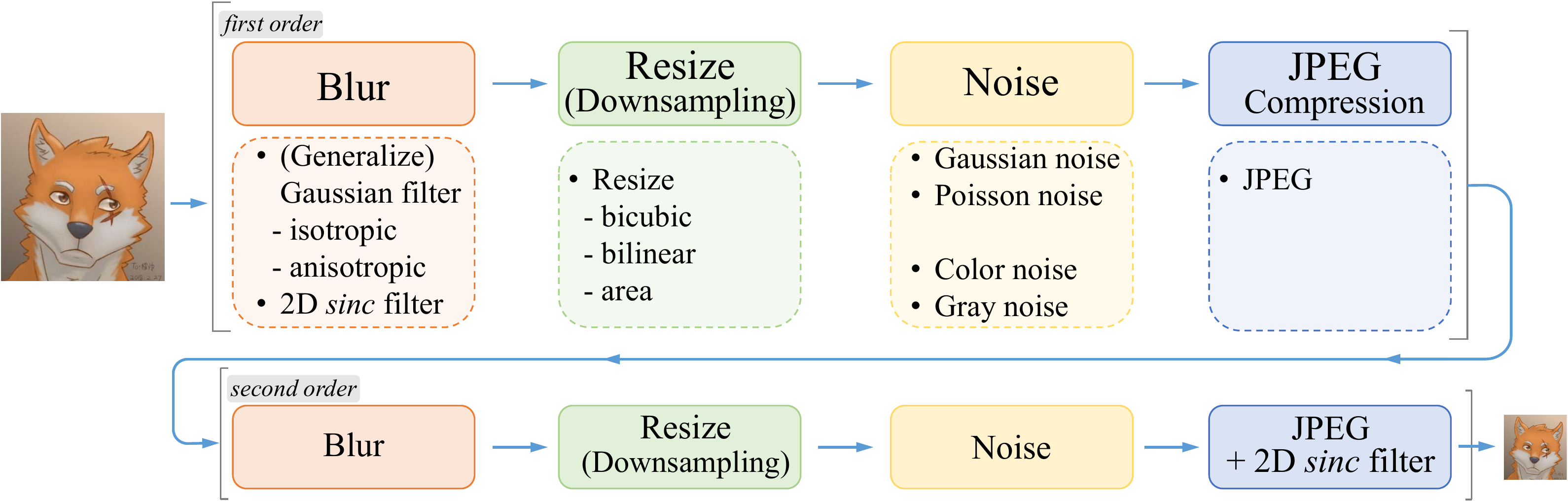}
	\end{center}
	\vspace{-0.6cm}
	\caption{Overview of the pure synthetic data generation adopted in Real-ESRGAN. It utilizes a second-order degradation process to model more practical degradations, where each degradation process adopts the classical degradation model.
	The detailed choices for blur , resize, noise and JPEG compression are listed. We also employ $sinc$ filter to synthesize common ringing and overshoot artifacts.}
	\label{fig:overview_degradation}
	\vspace{-0.5cm}
\end{figure*}

\begin{figure}[h]
	\vspace{-0.2cm}
	\begin{center}
		\includegraphics[width=\linewidth]{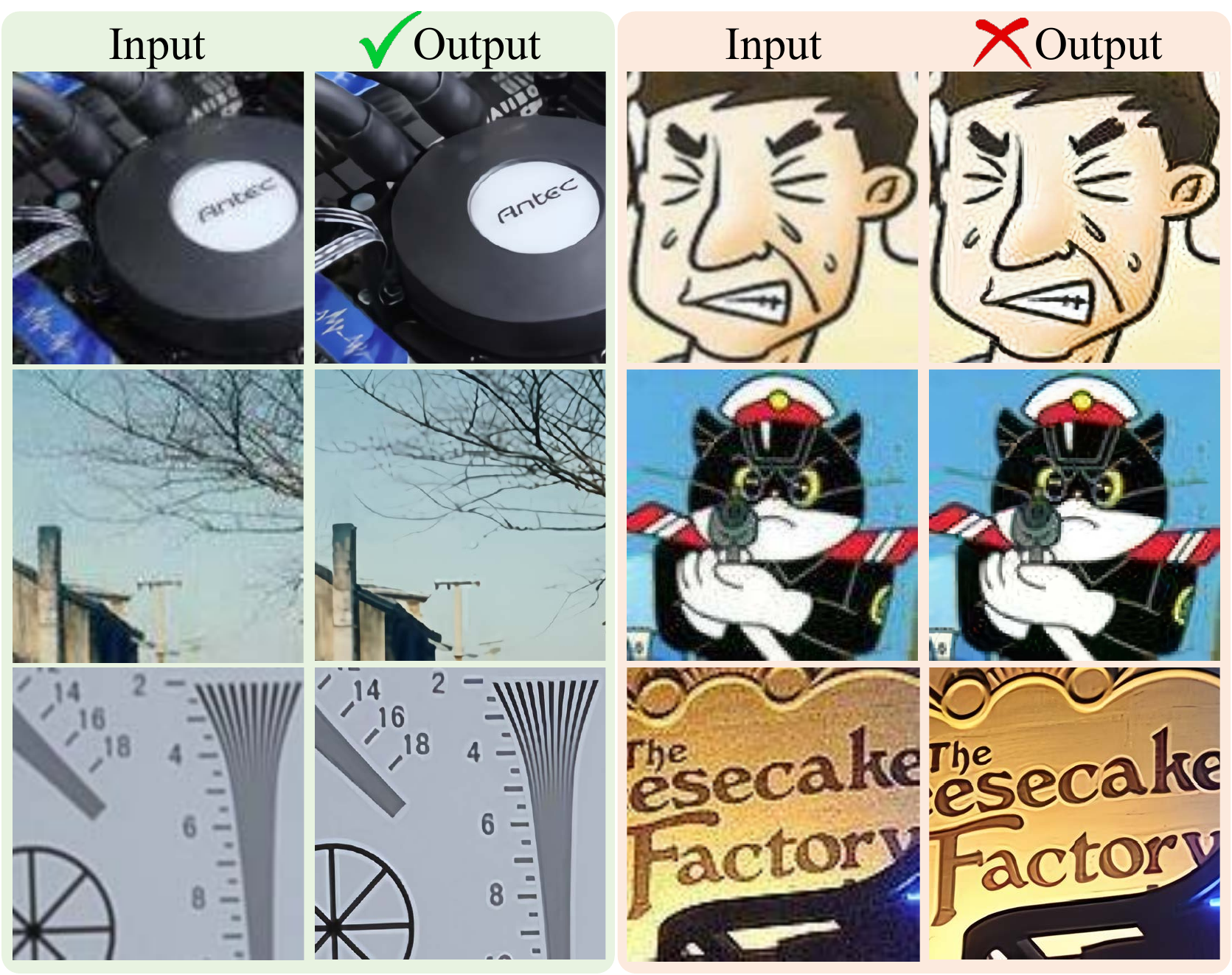}
	\end{center}
	\vspace{-0.6cm}
	\caption{Models trained with synthetic data of classical degradation model could resolve some real samples (Left). Yet, they amplify noises or introduce ringing artifacts for complex real-world images (Right). Zoom in for best view}
	\label{fig:classic_degradation_rlt}
	\vspace{-0.6cm}
\end{figure}

\subsection{High-order Degradation Model}
When we adopt the above classical degradation model to synthesize training pairs, the trained model could indeed handle some real samples. However, it still can not resolve some complicated degradations in the real world, especially the unknown noises and complex artifacts (see Fig.~\ref{fig:classic_degradation_rlt}).
It is because that the synthetic low-resolution images still have a large gap with realistic degraded images.
We thus extend the classical degradation model to a high-order degradation process to model more practical degradations.

The classical degradation model only includes a fixed number of  basic degradations, which can be regarded as a first-order modeling.
However, the real-life degradation processes are quite diverse, and usually comprise a series of procedures including imaging system of cameras, image editing, Internet transmission, \etc.
For instance, when we want to restore a low-quality image download from the Internet, its underlying degradation involves
a complicated combination of different degradation processes.
Specifically, the original image might be taken with a cellphone many years ago, which inevitably contains degradations such as camera blur, sensor noise, low resolution and JPEG compression.
The image was then edited with sharpening and resize operations, bringing in overshoot and blur artifacts.
After that, it was uploaded to some social media applications, which introduces a further compression and unpredictable noises. As the digital transmission will also bring artifacts, this process becomes more complicated when the image spreads several times on the Internet.
\begin{figure*}[t]
	\vspace{-0.6cm}
	\begin{center}
		\includegraphics[width=\linewidth]{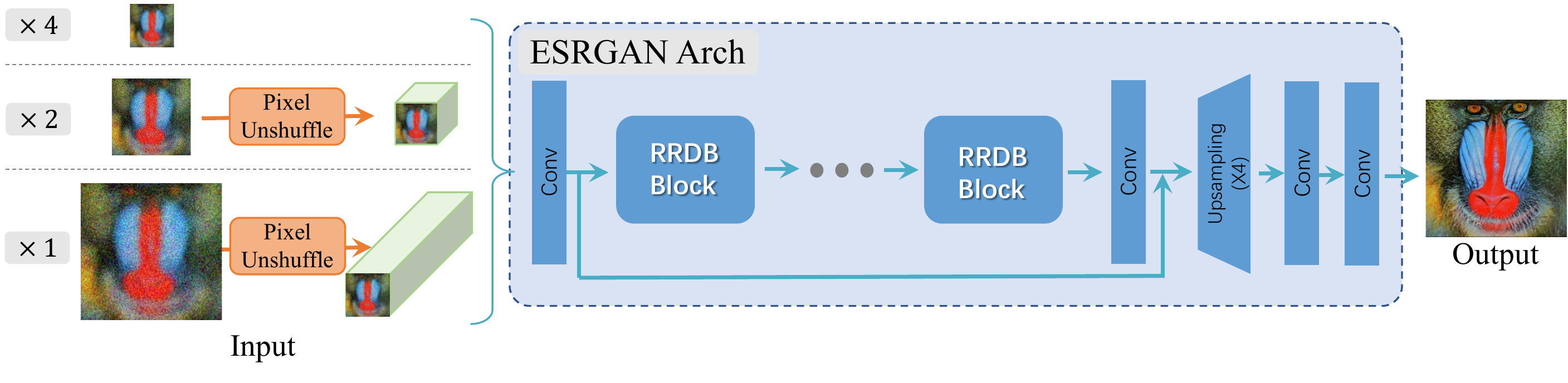}
	\end{center}
	\vspace{-0.8cm}
	\caption{Real-ESRGAN adopts the same generator network as that in ESRGAN. For the scale factor of $\times 2$ and $\times 1$, it first employs a pixel-unshuffle operation to reduce spatial size and re-arrange information to the channel dimension.}
	\label{fig:arch}
	\vspace{-0.3cm}
\end{figure*}
\begin{figure}[h]
	\begin{center}
		\includegraphics[width=\linewidth]{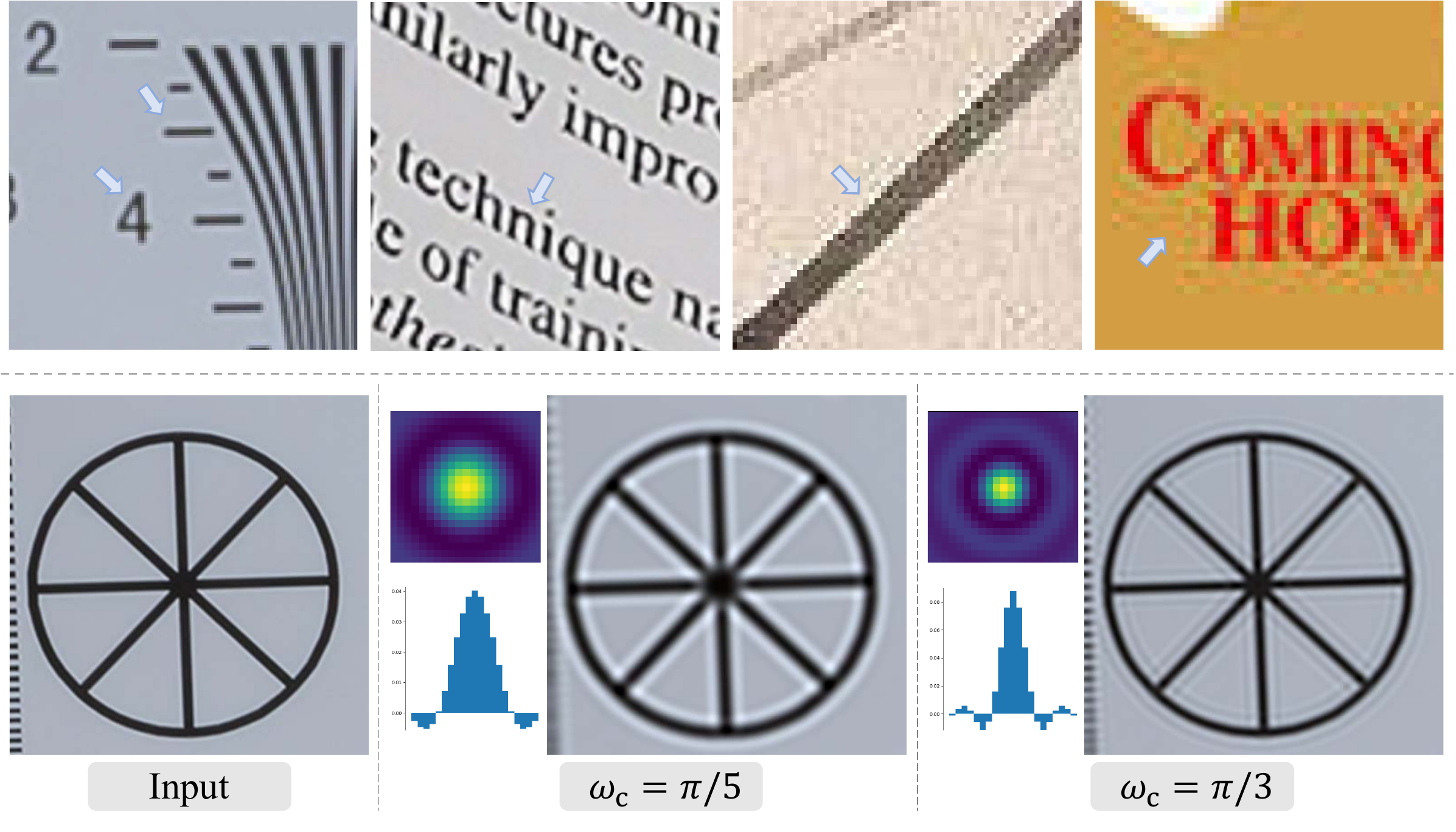}
	\end{center}
	\vspace{-0.5cm}
	\caption{\textbf{Top}: Real samples suffering from ringing and overshoot artifacts. \textbf{Bottom}: Examples of $sinc$ kernels (kernel size 21) and the corresponding filtered images. Zoom in for best view}
	\label{fig:sinc_samples}
	\vspace{-0.3cm}
\end{figure}

Such a complicated deterioration process could not be modeled with the classical first-order model. Thus, we propose a high-order degradation model.  An $n$-order model involves $n$ repeated degradation processes (as shown in Eq.~\ref{equ:n_degradation}), where each degradation process adopts the classical degradation model (Eq.~\ref{equ:degradation}) with the same procedure but different hyper-parameters.
Note that the ``high-order'' here is different from that used in mathematical functions. It mainly refers to the implementation time of the same operation.
The random shuffling strategy in~\cite{zhang2021designing} may also include repeated degradation processes (\eg, double blur or JPEG).
But we highlight that the high-order degradation process is the key, indicating that not all the shuffled degradations are necessary.
In order to keep the image resolution in a reasonable range, the downsampling operation in Eq.~\ref{equ:degradation} is replaced with a random resize operation.
Empirically, we adopt a second-order degradation process, as it could resolve most real cases while keeping simplicity.
Fig.~\ref{fig:overview_degradation} depicts the overall pipeline of our pure synthetic data generation pipeline.

\begin{equation}\label{equ:n_degradation}
	\bm{x} = \mathcal{D}^n(\bm{y}) = (\mathcal{D}_n \circ \cdots\circ \mathcal{D}_2\circ \mathcal{D}_1)(\bm{y}).
\end{equation}
It is worth noting that the improved high-order degradation process is not perfect and could not cover the whole degradation space in the real world. Instead, it merely extends the solvable degradation boundary of previous blind SR methods through modifying the data synthesis process. Several typical limitation scenarios can be found in Fig.~\ref{fig:limitation}.

\subsection{Ringing and overshoot artifacts}
Ringing artifacts often appear as spurious edges near sharp transitions in an image. They visually look like bands or ``ghosts'' near edges.
Overshoot artifacts are usually combined with ringing artifacts, which manifest themselves as an increased jump at the edge transition.
The main cause of these artifacts is that the signal is bandlimited without high frequencies.
These artifacts are very common and usually produced by a sharping algorithm, JPEG compression, \etc.
Fig.~\ref{fig:sinc_samples} (Top) shows some real samples suffering from ringing and overshoot artifacts.

We employ the $sinc$ filter, an idealized filter that cuts off high frequencies, to synthesize ringing and overshoot artifacts for training pairs.
The $sinc$ filter kernel can be expressed as\footnote{We use the implementation in \href{https://dsp.stackexchange.com/questions/58301/2-d-circularly-symmetric-low-pass-filter}{this url}.}:
\begin{align}\label{equ:sinc}
	\bm{k}(i,j) &= \frac{\omega_c}{2\pi\sqrt{i^2+j^2}} J_1(\omega_c \sqrt{i^2+j^2}),
\end{align}
where $(i, j)$ is the kernel coordinate; $\omega_c$ is the cutoff frequency; and $J_1$ is
the first order Bessel function of the first kind.
Fig.~\ref{fig:sinc_samples} (Bottom) shows $sinc$ filters with different cutoff frequencies, and their corresponding filtered images. It is observed that it could well synthesize ringing and overshoot artifacts (especially introduced by over-sharp effects). These artifacts are visually similar to those in the first two real samples in Fig.~\ref{fig:sinc_samples} (Top).

\begin{figure}[t]
	\begin{center}
		\includegraphics[width=\linewidth]{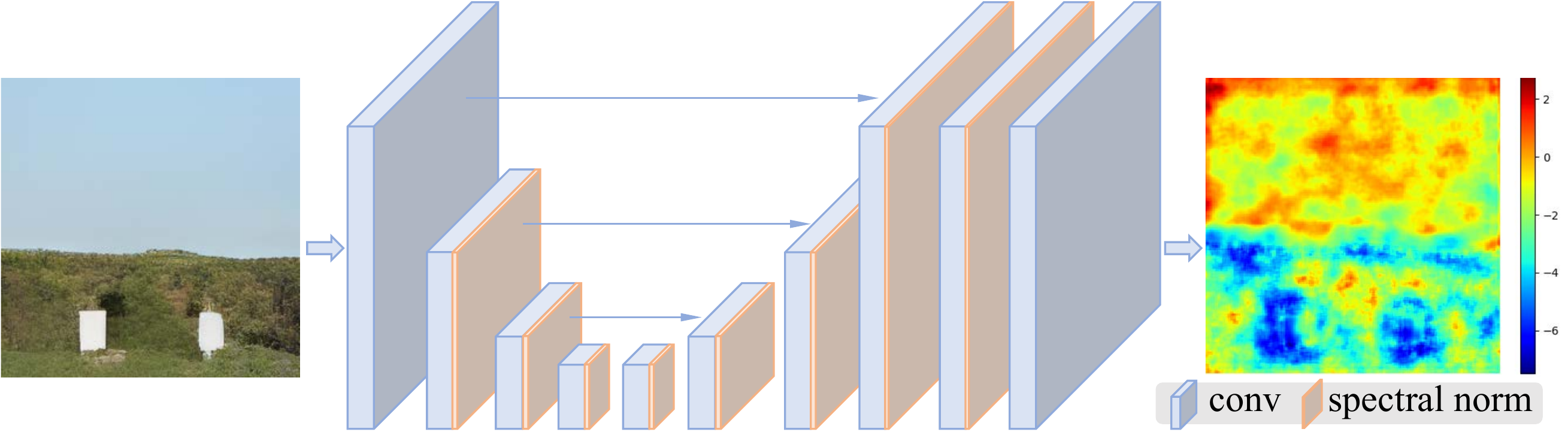}
	\end{center}
	\vspace{-0.5cm}
	\caption{Architecture of the U-Net discriminator with spectral normalization.}
	\label{fig:unet_disc}
	\vspace{-0.5cm}
\end{figure}

We adopt $sinc$ filters in two places: the blurring process and the last step of the synthesis.
The order of the last $sinc$ filter and JPEG compression is randomly exchanged to cover a larger degradation space, as some images may be first over-sharpened (with overshoot artifacts) and then have JPEG compression; while some images may first do JPEG compression followed by sharpening operation.

\subsection{Networks and Training}
\noindent\textbf{ESRGAN generator}. We adopt the same generator (SR network) as ESRGAN~\cite{wang2018esrgan}, \ie, a deep network with several residual-in-residual dense blocks (RRDB), as shown in Fig.~\ref{fig:arch}.
We also extend the original $\times 4$ ESRGAN architecture to perform super-resolution with a scale factor of $\times 2$ and $\times 1$. As ESRGAN is a heavy network, we first employ the pixel-unshuffle (an inverse operation of pixel-shuffle~\cite{shi2016real}) to reduce the spatial size and enlarge the channel size before feeding inputs into the main ESRGAN architecture. Thus, the most calculation is performed in a smaller resolution space, which can reduce the GPU memory and computational resources consumption.

\noindent\textbf{U-Net discriminator with spectral normalization (SN).}
As Real-ESRGAN aims to address a much larger degradation space than ESRGAN, the original design of discriminator in ESRGAN is no longer suitable.
Specifically, the discriminator in Real-ESRGAN requires a greater discriminative power for complex training outputs.
Instead of discriminating global styles, it also needs to produce accurate gradient feedback for local textures.
Inspired by~\cite{schonfeld2020u,yan2021fine}, we also improve the VGG-style discriminator in ESRGAN to an U-Net design with skip connections (Fig.~\ref{fig:unet_disc}). The U-Net outputs realness values for each pixel, and can provide detailed per-pixel feedback to the generator.

In the meanwhile, the U-Net structure and complicate degradations also increase the training instability. We employ the spectral normalization regularization ~\cite{miyato2018spectral} to stabilize the training dynamics. Moreover, we observe that spectral normalization is also beneficial to alleviate the over-sharp and annoying artifacts introduced by GAN training.
With those adjustments, we are able to easily train the Real-ESRGAN and achieve a good balance of local detail enhancement and artifact suppression.

\noindent\textbf{The training process} is divided into two stages. First, we train a PSNR-oriented model with the L1 loss.
The obtained model is named by \textit{Real-ESRNet}.
We then use the trained PSNR-oriented model as an initialization of the generator, and train the \textit{Real-ESRGAN} with a combination of  L1 loss, perceptual loss~\cite{johnson2016perceptual} and GAN loss~\cite{goodfellow2014generative,ledig2017srgan,blau2018perception}.

%% file: sections/4_experiments.tex
\section{Experiments} \label{sec:experiments}

\subsection{Datasets and Implementation}
\label{sec:implementation}
\noindent\textbf{Training details.}
Similar to ESRGAN, we adopt DIV2K~\cite{agustsson2017ntire}, Flickr2K~\cite{timofte2017ntire} and OutdoorSceneTraining~\cite{wang2018sftgan} datasets for training.
The training HR patch size is set to 256. We train our models with four NVIDIA V100 GPUs with a total batch size of 48.
We employ Adam optimizer~\cite{kingma2014adam}. Real-ESRNet is finetuned from ESRGAN for faster convergence. We train Real-ESRNet for $1000K$ iterations with learning rate $2\times 10^{-4}$ while training Real-ESRGAN for $400K$ iterations with learning rate $1\times 10^{-4}$.
We adopt exponential moving average (EMA) for more stable training and better performance.
Real-ESRGAN is trained with a combination of L1 loss, perceptual loss and GAN loss, with weights $\{1, 1, 0.1\}$, respectively.
We use the $\{\mathtt{conv1}, ... \mathtt{conv5}\}$  feature maps (with weights $\{0.1, 0.1, 1, 1, 1\}$) before activation in the pre-trained VGG19 network~\cite{johnson2016perceptual} as the perceptual loss. Our implementation is based on the BasicSR~\cite{wang2020basicsr}.

\noindent\textbf{Degradation details.}
We employ a second-order degradation model for a good balance of simplicity and effectiveness.
Unless otherwise specified, the two degradation processes have the same settings.
 We adopt Gaussian kernels, generalized Gaussian kernels and plateau-shaped kernels, with a probability of $\{0.7, 0.15, 0.15\}$. The blur kernel size is randomly selected from $\{7, 9, ... 21$\}. Blur standard deviation $\sigma$ is sampled from $[0.2, 3]$ ($[0.2, 1.5]$ for the second degradation process). Shape parameter $\beta$ is sampled from $[0.5, 4]$ and $[1, 2]$ for generalized Gaussian and plateau-shaped kernels, respectively. We also use $sinc$ kernel with a probability of 0.1. We skip the second blur degradation with a probability of 0.2.

We employ Gaussian noises and Poisson noises with a probability of $\{0.5, 0.5\}$. The noise sigma range and Poisson noise scale are set to $[1, 30]$ and $[0.05, 3]$, respectively ($[1, 25]$ and $[0.05, 2.5]$ for the second degradation process).
The gray noise probability is set to 0.4.
JPEG compression quality factor is set to $[30, 95]$.
The final $sinc$ filter is applied with a probability of 0.8.
More details can be found in the \href{https://github.com/xinntao/Real-ESRGAN}{released codes}.

\noindent\textbf{Training pair pool.}
In order to improve the training efficiency, all degradation processes are implemented in PyTorch with CUDA acceleration, so that we are able to synthesize training pairs on the fly.
However, batch processing limits the diversity of synthetic degradations in a batch. For example, samples in a batch could not have different resize scaling factors.
Therefore, we employ a training pair pool to increase the degradation diversity in a batch.
At each iteration, the training samples are randomly selected from the training pair poor to form a training batch.
We set the pool size to 180 in our implementation.

\noindent\textbf{Sharpen ground-truth images during training}.
We further show a training trick to visually improve the sharpness, while not introducing visible artifacts.
A typical way of sharpening images is to employ a post-process algorithm, such as unsharp masking (USM).
However, this algorithm tends to introduce overshoot artifacts.
We empirically find that sharpening ground-truth images during training could achieve a better balance of sharpness and overshoot artifact suppression.
We denote the model trained with sharped ground-truth images as Real-ESRGAN\textbf{+} (comparisons are shown in Fig.~\ref{fig:compare}).

\subsection{Comparisons with Prior Works}
We compare our Real-ESRGAN with several state-of-the-art methods, including ESRGAN~\cite{wang2018esrgan}, DAN~\cite{luo2020unfolding}, CDC~\cite{wei2020cdc}, RealSR~\cite{ji2020real} and BSRGAN~\cite{zhang2021designing}.
We test on several diverse testing datasets with real-world images, including RealSR~\cite{cai2019toward}, DRealSR~\cite{wei2020cdc}, OST300~\cite{wang2018sftgan}, DPED~\cite{ignatov2017dslr}, ADE20K validation~\cite{zhou2019semantic} and images from Internet.
Since existing metrics for perceptual quality cannot well reflect the actual human perceptual preferences on the fine-grained scale~\cite{blau20182018}, we present several representative visual samples in Fig.~\ref{fig:compare}.
The quantitative results are also included in the Appendix.~\ref{sec:quantitative} for reference.

It can be observed from~Fig.~\ref{fig:compare} that our Real-ESRGAN outperforms previous approaches in both removing artifacts and restoring texture details. Real-ESRGAN+ (trained with sharpened ground-truths) can further boost visual sharpness.
Specifically, the first sample contains overshoot artifacts (white edges around letters). Directly upsampling will inevitably amplify those artifacts (\eg, DAN and BSRGAN). Real-ESRGAN takes such common artifacts into consideration and simulates them with $sinc$ filter, thus effectively removing ringing and overshoot artifacts.
The second sample contains unknown and complicated degradations. Most algorithms can not effectively eliminate them while Real-ESRGAN trained with second-order degradation processes could.
Real-ESRGAN is also capable of restoring more realistic textures (\eg, brick, mountain and tree textures) for real-world samples, while other methods either fail to remove degradations or add unnatural textures (\eg, RealSR and BSRGAN).

\begin{figure*}
	\vspace{-0.2cm}
	\begin{center}
		\includegraphics[width=1\linewidth]{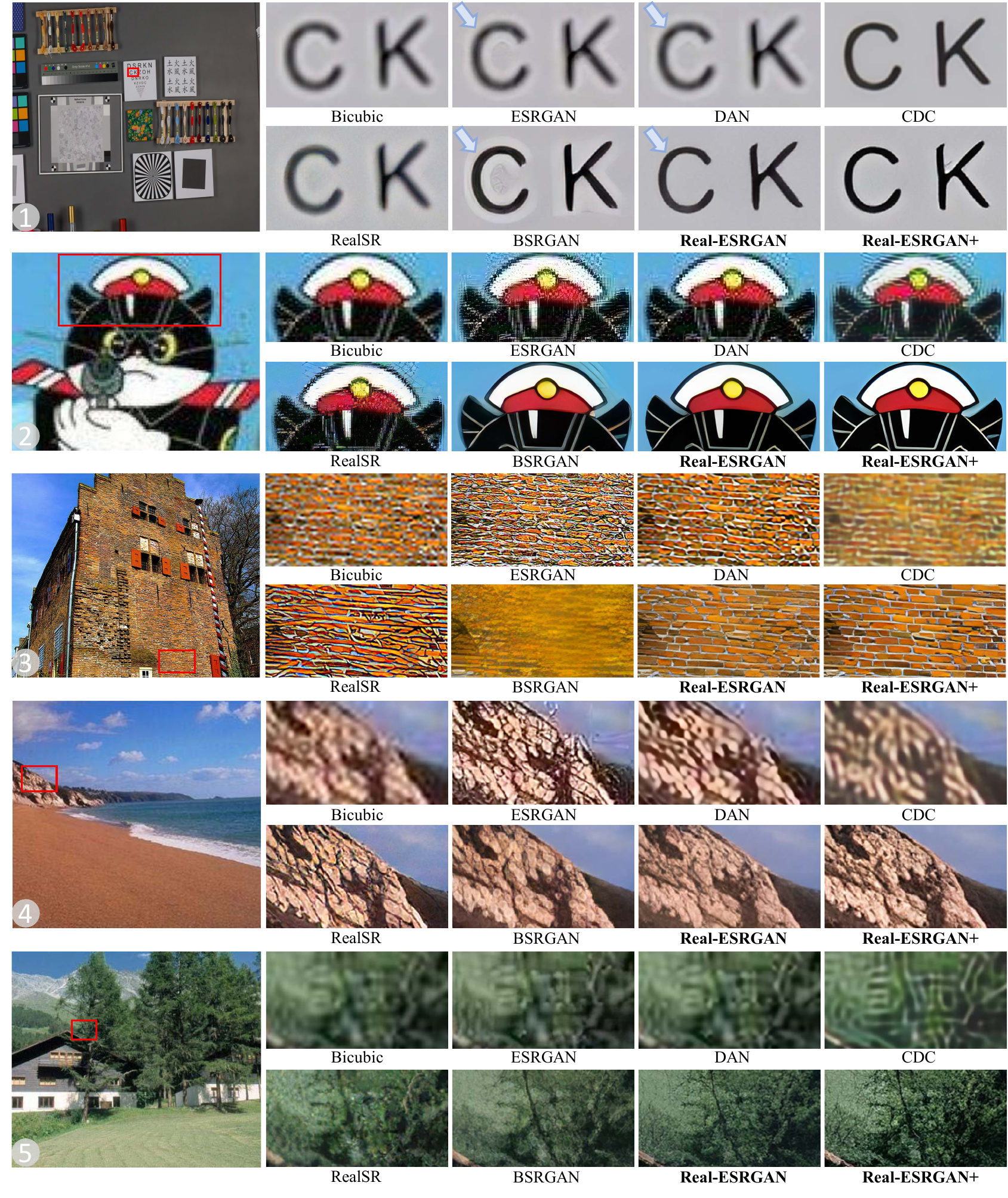}
	\end{center}
	\vspace{-0.5cm}
	\caption{Qualitative comparisons on several representative real-world samples with upsampling scale factor of 4. Our \mbox{Real-ESRGAN} outperforms previous approaches in both removing artifacts and restoring texture details. Real-ESRGAN+ (trained with sharpened ground-truths) can further boost visual sharpness.
	Other methods may either fail to remove overshoot (the 1st sample) and complicated artifacts (the 2nd sample), or fail to restore realistic and natural textures for various scenes (the 3rd, 4th, 5th samples).
	(\textbf{Zoom in for best view})}
	\label{fig:compare}
\end{figure*}

\subsection{Ablation Studies}
\begin{figure}[t]
	\vspace{-0.2cm}
	\begin{center}
		\includegraphics[width=1\linewidth]{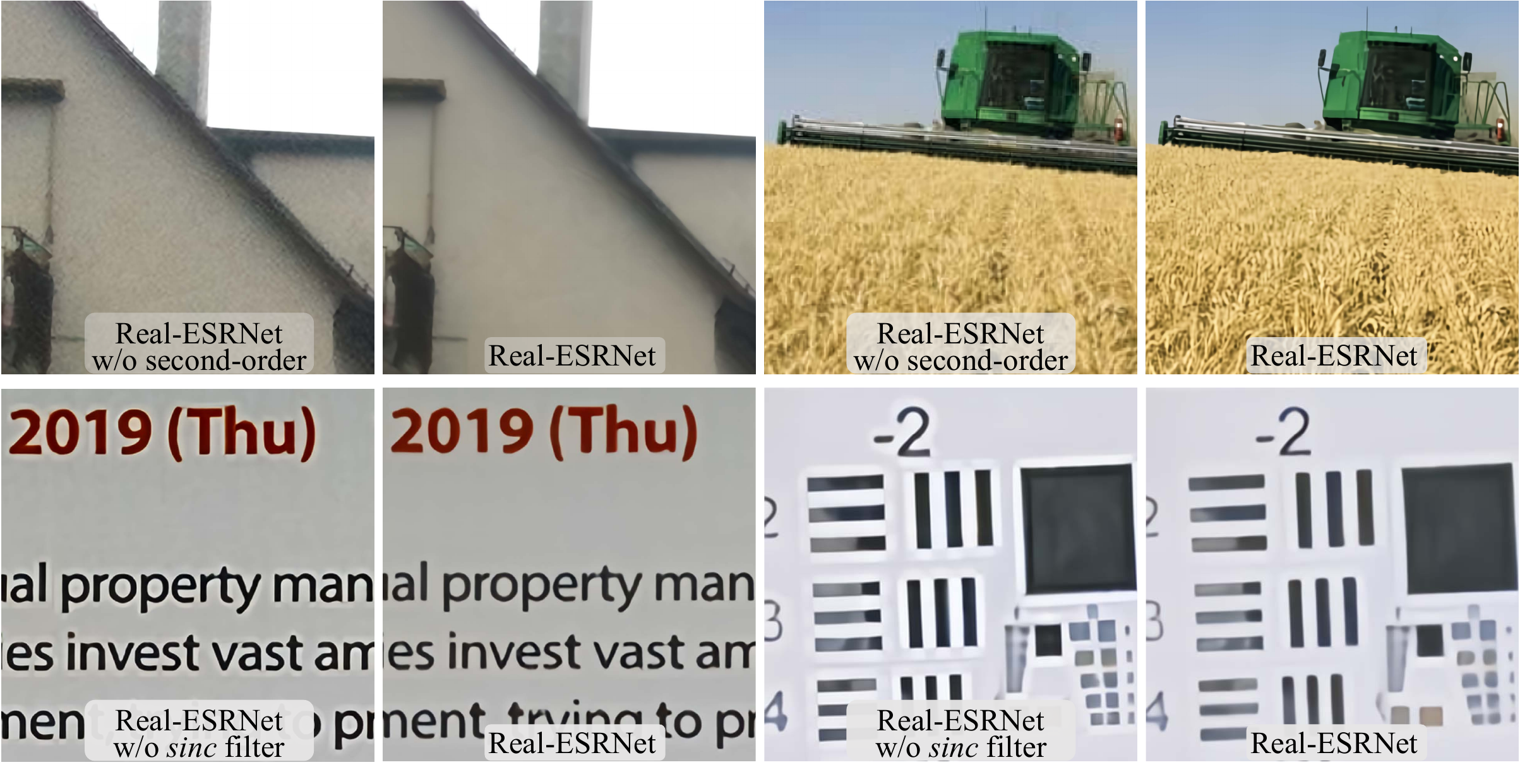}
	\end{center}
	\vspace{-0.6cm}
	\caption{\textbf{Top}: Real-ESRNet results w/ and w/o second-order degradation process. \textbf{Bottom}: Real-ESRNet results w/ and w/o $sinc$ filters. Zoom in for best view}
	\label{fig:ablation_secondorder_sinc}
	\vspace{-0.5cm}
\end{figure}

\noindent\textbf{Second-order degradation model}.
We conduct ablation studies of degradations on Real-ESRNet, as it is more controllable and can better reflect the influence of degradations.
We replace the second-order process in Real-ESRNet with the classical degradation model to generate training pairs.
As shown in Fig.~\ref{fig:ablation_secondorder_sinc} (Top), models trained with classical first-order degradation model cannot effectively remove noise on the wall or blur in the wheat field, while Real-ESRNet can handle these cases.

\noindent\textbf{\textit{sinc} filters}.
If $sinc$ filters are not employed during training, the restored results will amplify the ringing and overshoot artifacts that existed in the input images, as shown in Fig.~\ref{fig:ablation_secondorder_sinc} (Bottom), especially around the text and lines. In contrast, models trained with $sinc$ filters can remove those artifacts.

\begin{figure}[t]
	\vspace{-0.2cm}
	\begin{center}
		\includegraphics[width=1\linewidth]{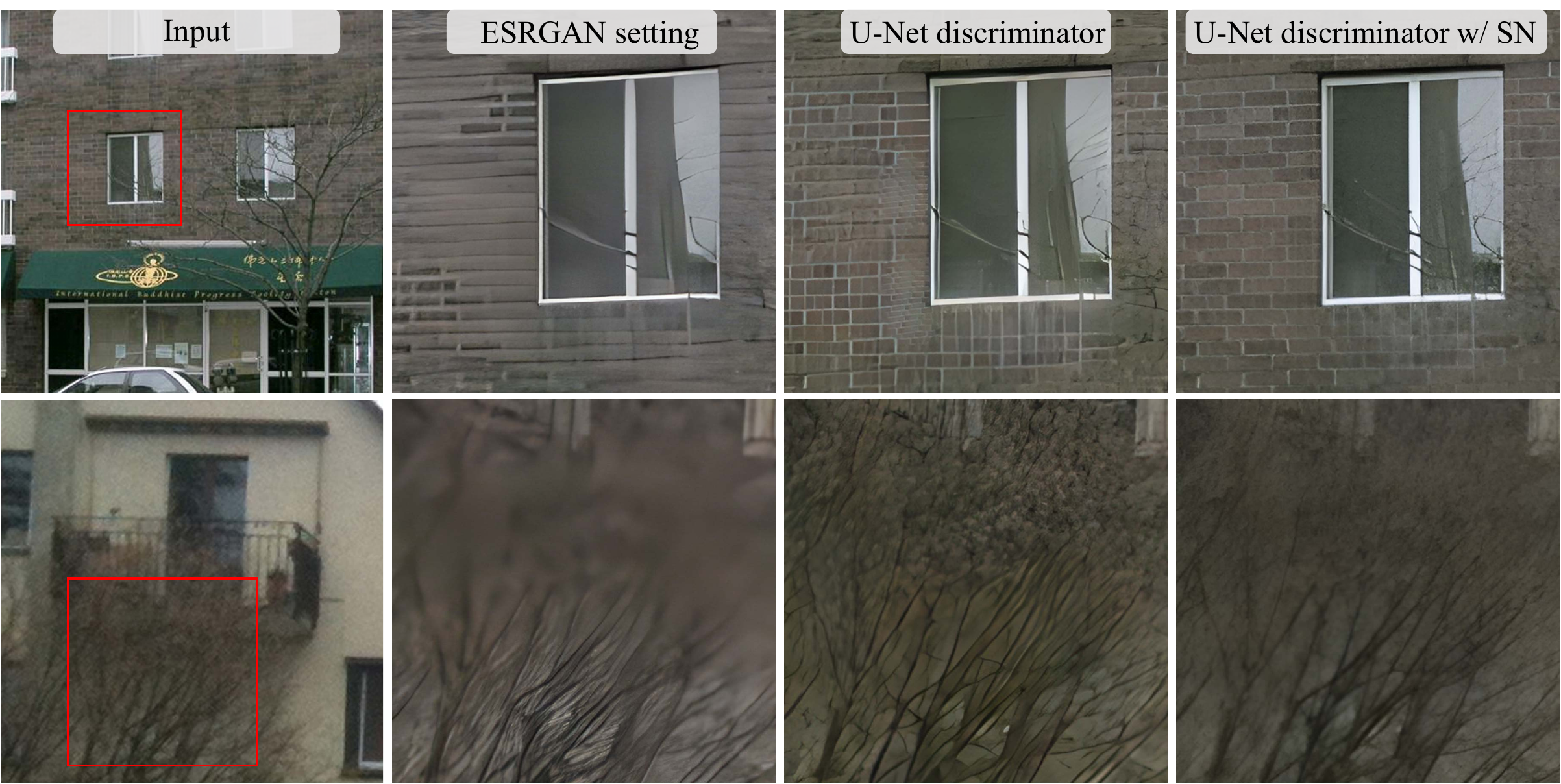}
	\end{center}
	\vspace{-0.7cm}
	\caption{Ablation on the discriminator design. Zoom in for best view}
	\label{fig:ablation_discriminator}
	\vspace{-0.2cm}
\end{figure}

\noindent\textbf{U-Net discriminator with SN regularization}.
We first employ the ESRGAN setting including the VGG-style discriminator and its loss weights. However, we can observe from Fig.~\ref{fig:ablation_discriminator}, this model cannot restore detailed textures (bricks and bushes) and even brings unpleasant artifacts in bush branches. Using a U-Net design could improve local details. Yet, it introduces unnatural textures and also increases training instability. SN regularization could improve restored textures while stabilizing training dynamics.

\begin{figure}[t]
	\vspace{-0.2cm}
	\begin{center}
		\includegraphics[width=1\linewidth]{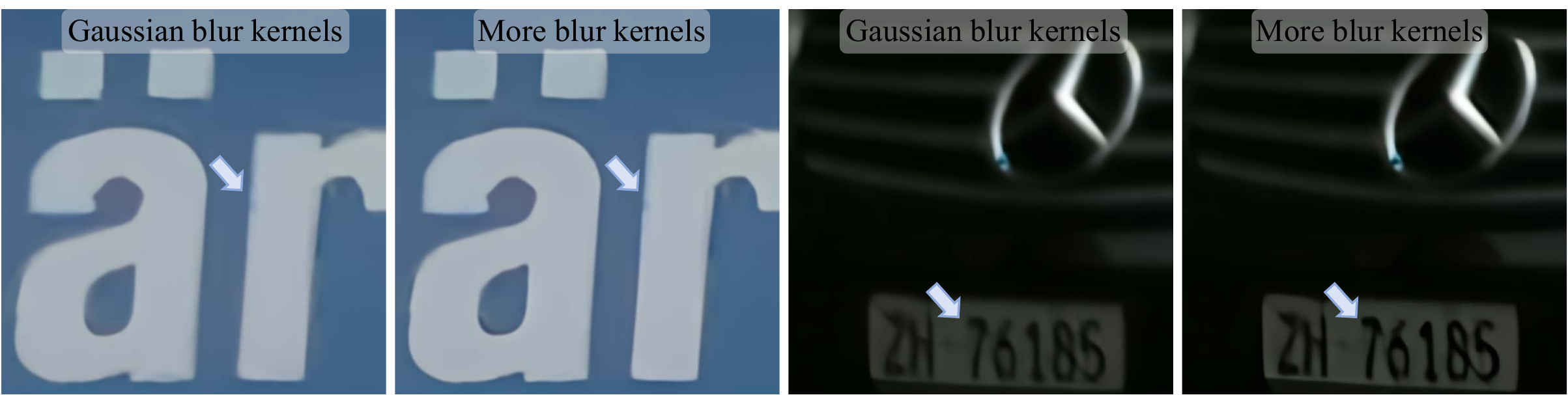}
	\end{center}
	\vspace{-0.7cm}
	\caption{Ablation on using more blur kernels (generalized blur and plateau-shaped kernels). Zoom in for best view}
	\label{fig:ablation_blurkernel}
\end{figure}

\noindent\textbf{More complicated blur kernels}.
We remove the generalized Gaussian kernel and plateau-shaped kernel in blur synthesis. As shown in Fig.~\ref{fig:ablation_blurkernel}, on some real samples, the model cannot remove blur and recover sharp edges as Real-ESRGAN do. Nevertheless, on most samples, their differences are marginal, indicating that the widely-used Gaussian kernels with a high-order degradation process can already cover a large real blur space. As we can still observe slightly better performance, we adopt those more complicated blur kernels in Real-ESRGAN.

\subsection{Limitations}
\begin{figure}[t]
	\vspace{-0.2cm}
	\begin{center}
		\includegraphics[width=\linewidth]{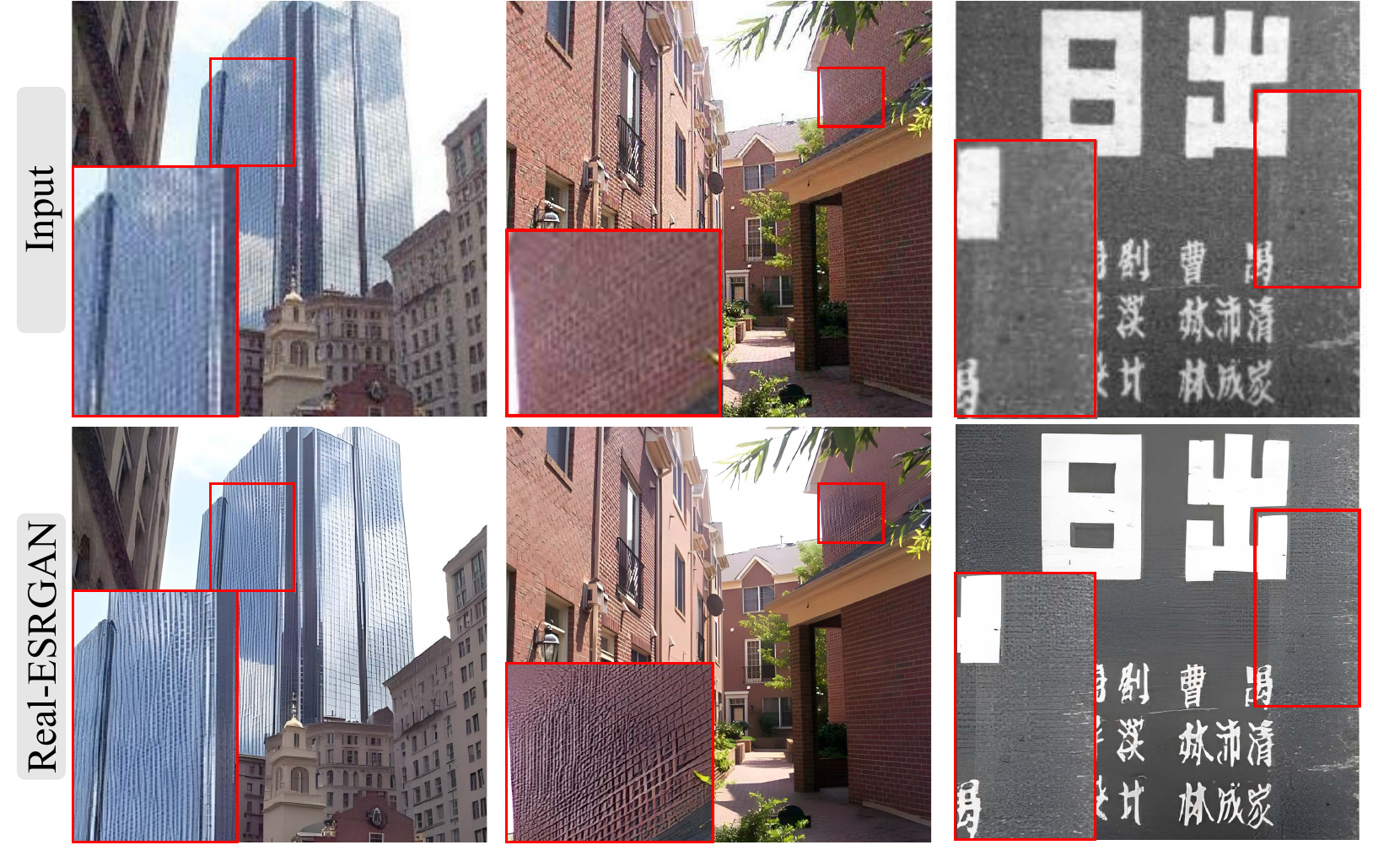}
	\end{center}
	\vspace{-0.6cm}
	\caption{Limitations: 1) twisted lines; 2) unpleasant artifacts caused by GAN training; 3) unknown and out-of-distribution degradations. Zoom in for best view}
	\label{fig:limitation}
	\vspace{-0.5cm}
\end{figure}

\vspace{-0.2cm}
Though Real-ESRGAN is able to restore most real-world images, it still has some limitations.
As shown in Fig.~\ref{fig:limitation},
1) some restored images (especially building and indoor scenes) have twisted lines due to aliasing issues.
2) GAN training introduces unpleasant artifacts on some samples.
3) It could not remove out-of-distribution complicated degradations in the real world. Even worse, it may amplify these artifacts.
These drawbacks have great impact on the practical application of Real-ESRGAN, which are in urgent need to address in future works.

%% file: sections/5_appendix.tex

\noindent\textbf{\large{Appendix}} \label{appendix}

\section{Details of Classical Degradation Model} \label{sec:more_detail_degradation_model}

In this section, we provide more details (especially examples) of each degradation type used in the classical degradation model.

\subsection{Blur}
\begin{figure}[h]
	\begin{center}
		\includegraphics[width=\linewidth]{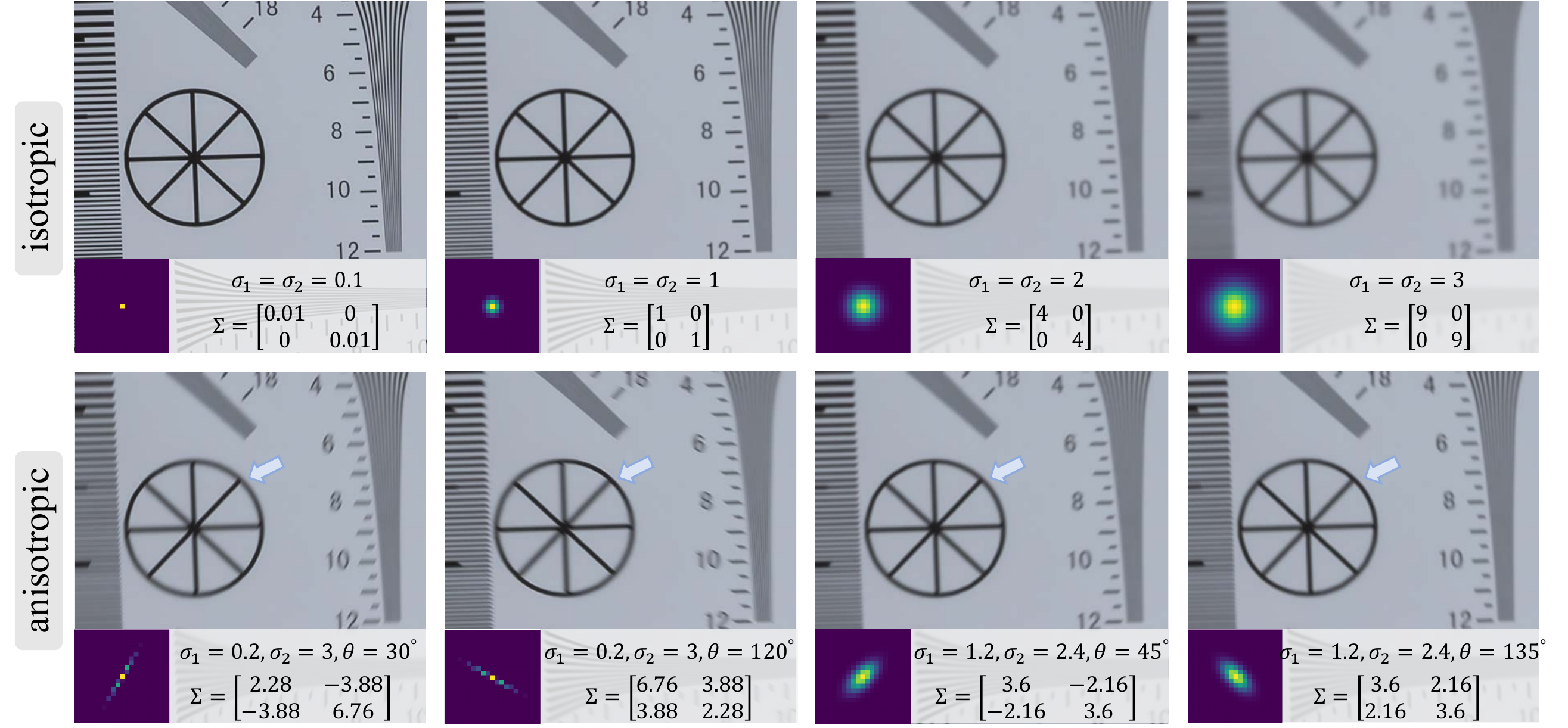}
	\end{center}
	\vspace{-0.5cm}
	\caption{Examples of Gaussian kernels (kernel size 21) and their corresponding blurry images. Zoom in for best view}
	\label{fig:blur_kernel}
\end{figure}

Isotropic and anisotropic Gaussian filters are the common choices for blur kernels. We show several Gaussian kernels and their corresponding blurry images in Fig.~\ref{fig:blur_kernel}.

\begin{figure}[h]
	\begin{center}
		\includegraphics[width=\linewidth]{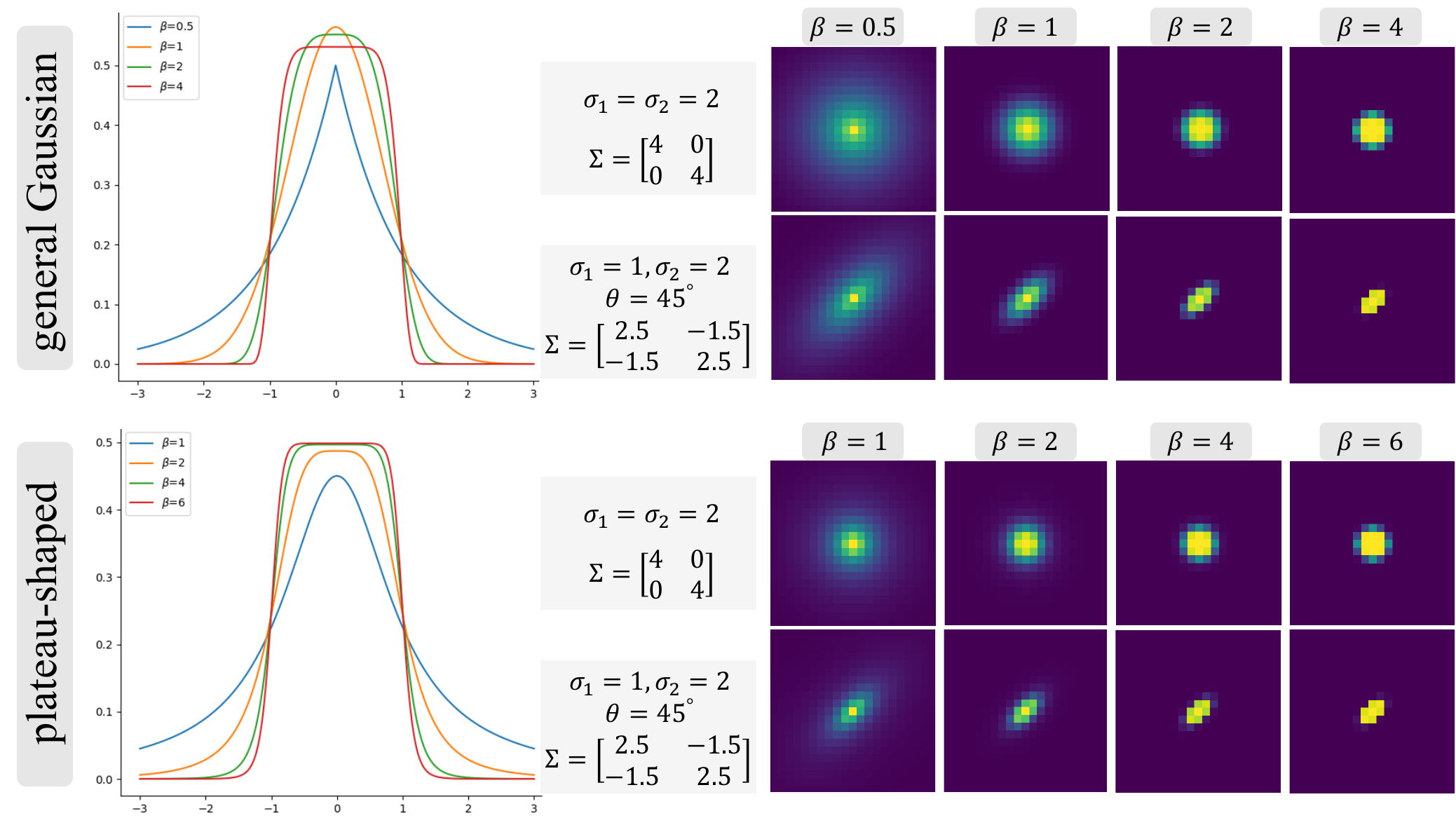}
	\end{center}
	\vspace{-0.5cm}
	\caption{Blur kernels with different shape parameters in general Gaussian distribution and plateau-shaped distribution. Zoom in for best view}
	\label{fig:blur_kernel_general}
\end{figure}

To include more diverse kernel shapes, we further adopt generalized Gaussian blur kernels~\cite{liu2020estimating} and a plateau-shaped distribution.
Fig.~\ref{fig:blur_kernel_general} shows how the shape parameter $\beta$ controls kernel shapes.
Empirically, we found that including these blur kernels produces sharper outputs for several real samples.

\subsection{Noise}
\begin{figure}[h]
	\begin{center}
		\includegraphics[width=\linewidth]{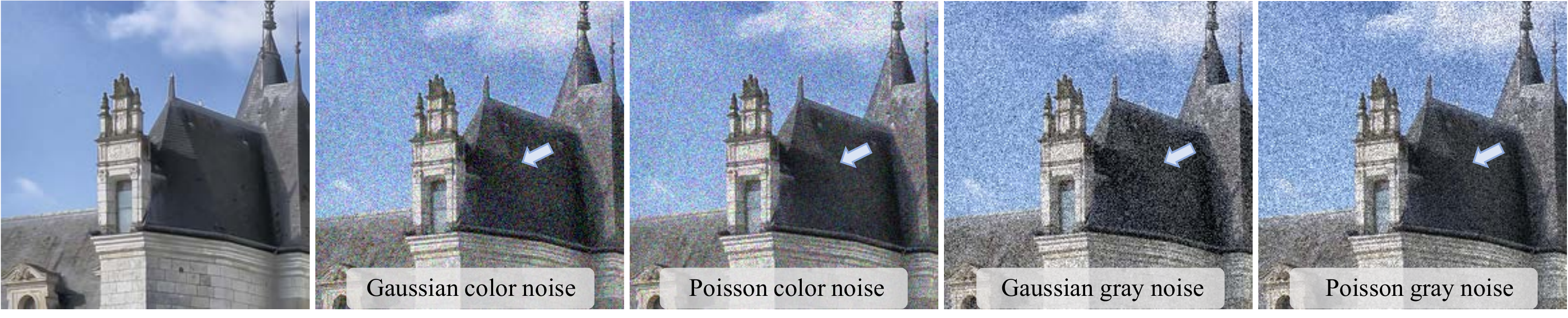}
	\end{center}
	\vspace{-0.5cm}
	\caption{Visual comparisons of Gaussian and Poisson noises. Poisson noise has low noise intensity in dark areas. Zoom in for best view}
	\label{fig:noise}
\end{figure}

Fig.~\ref{fig:noise} depicts the additive Gaussian noise and Poisson noise.
Poisson noise has an intensity proportional to the image intensity, and the noises at different pixels are independent of one another.
As shown in Fig.~\ref{fig:noise}, the Poisson noise has low noise intensity in dark areas.

\begin{table*}[]
	\caption{NIQE scores on several diverse testing datasets with real-world images. The lower, the better.}
	\label{tab:qualitative}
	\resizebox{\textwidth}{!}{%
		\begin{tabular}{l|llllllll}
			\hline
			& \multicolumn{1}{c}{Bicubic} & \multicolumn{1}{c}{ESRGAN~\cite{wang2018esrgan}} & \multicolumn{1}{c}{DAN~\cite{luo2020unfolding}} & \multicolumn{1}{c}{RealSR~\cite{ji2020real}} & \multicolumn{1}{c}{CDC~\cite{wei2020cdc}} & BSRGAN~\cite{zhang2021designing} & \multicolumn{1}{c}{\textbf{Real-ESRGAN}} & \multicolumn{1}{c}{\textbf{Real-ESRGAN+}} \\ \hline
			RealSR-Canon~\cite{cai2019toward} & 6.1269                      & 6.7715                     & 6.5282                  & 6.8692                     & 6.1488                  & 5.7489 & 4.5899                                   & \textbf{4.5314}                           \\
			RealSR-Nikon~\cite{cai2019toward} & 6.3607                      & 6.7480                     & 6.6063                  & 6.7390                     & 6.3265                  & 5.9920 & 5.0753                                   & \textbf{5.0247}                           \\
			DRealSR~\cite{wei2020cdc}      & 6.5766                      & 8.6335                     & 7.0720                  & 7.7213                     & 6.6359                  & 6.1362 & 4.9796                                   & \textbf{4.8458}                           \\
			DPED-iphone~\cite{ignatov2017dslr}  & 6.0121                      & 5.7363                     & 6.1414                  & 5.5855                     & 6.2738                  & 5.9906 & 5.4352                                   & \textbf{5.2631}                           \\
			OST300~\cite{wang2018sftgan}       & 4.4440       &3.5245                                   & 5.0232                  & 4.5715                     & 4.7441                  & 4.1662 & 2.8659                                   & \textbf{2.8191}                           \\
			ImageNet val~\cite{deng2009imagenet}  & 7.4985        &\textbf{3.6474}                                    & 6.0932              & 3.8303                      & 7.0441                   & 4.3528  & 4.8580                                   & 4.6448                            \\
			ADE20K val~\cite{zhou2019semantic}  & 7.5239         &3.6905                                     & 6.3839               & \textbf{3.4102}                      & 6.9219                   & 3.9434  & 3.7886                                   & 3.5778                            \\
			\hline
	\end{tabular}}
\end{table*}

\subsection{Resize}
\begin{figure}[h]
	\begin{center}
		\includegraphics[width=\linewidth]{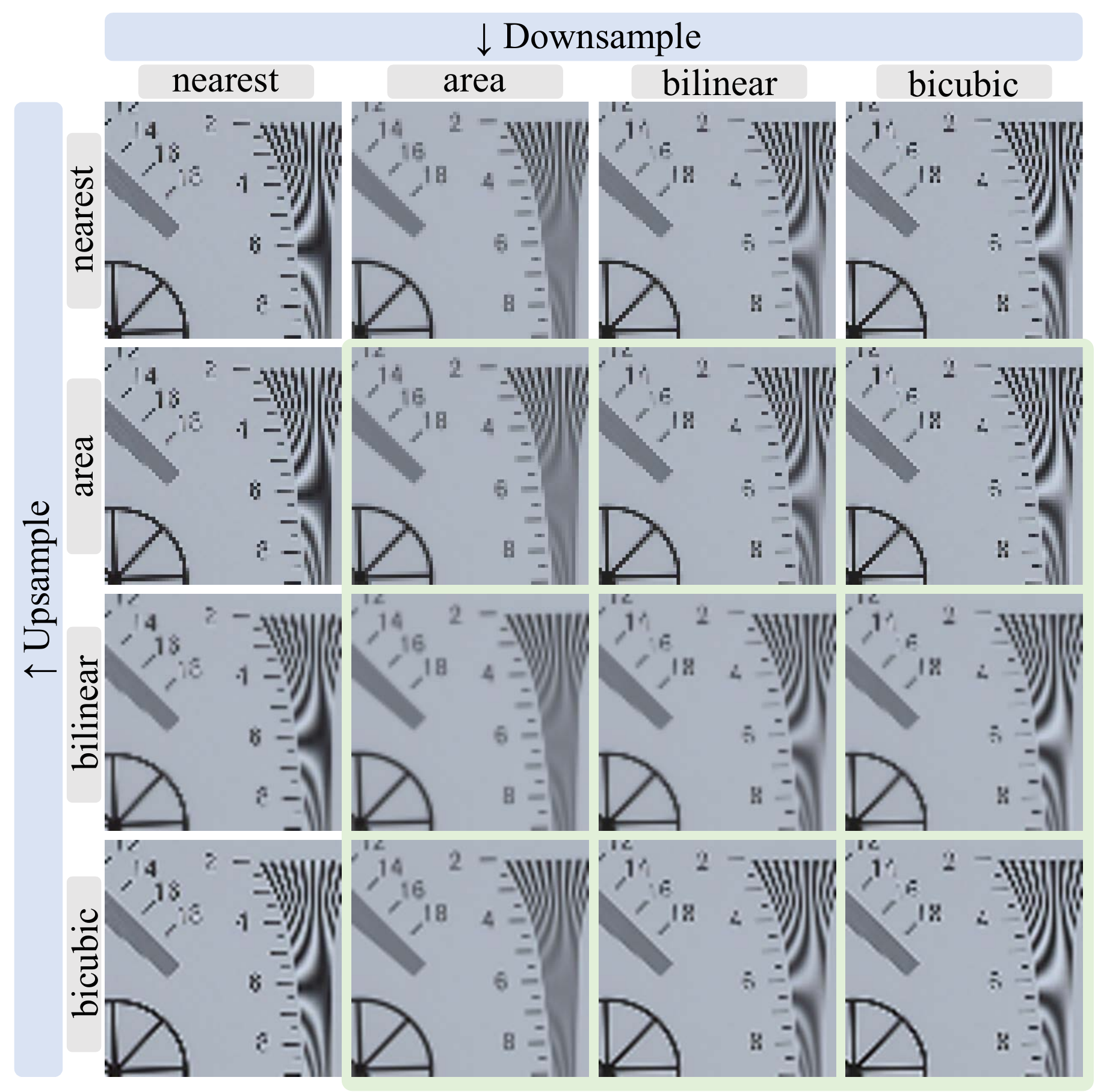}
	\end{center}
	\vspace{-0.5cm}
	\caption{Effects of different combinations of down- and up-sampling algorithms. The images are first downsampled by a scale factor of four and then upsampled to its original size. Zoom in for best view}
	\label{fig:resize}
\end{figure}

There are several resize algorithms.
We compare the following resize operations: nearest-neighbor interpolation, area resize, bilinear interpolation and bicubic interpolation.
We examine the different effects of these resize operations. We first downsample an image by a scale factor of four and then upsample to its original size. Different downsampling and upsampling algorithms are performed, and the results of different combinations are shown in Fig.~\ref{fig:resize}.
It is observed that different resize operations result in very different effects - some produce blurry results while some may output over-sharp images with overshoot artifacts.

\subsection{JPEG compression}
\begin{figure}[h]
	\begin{center}
		\includegraphics[width=\linewidth]{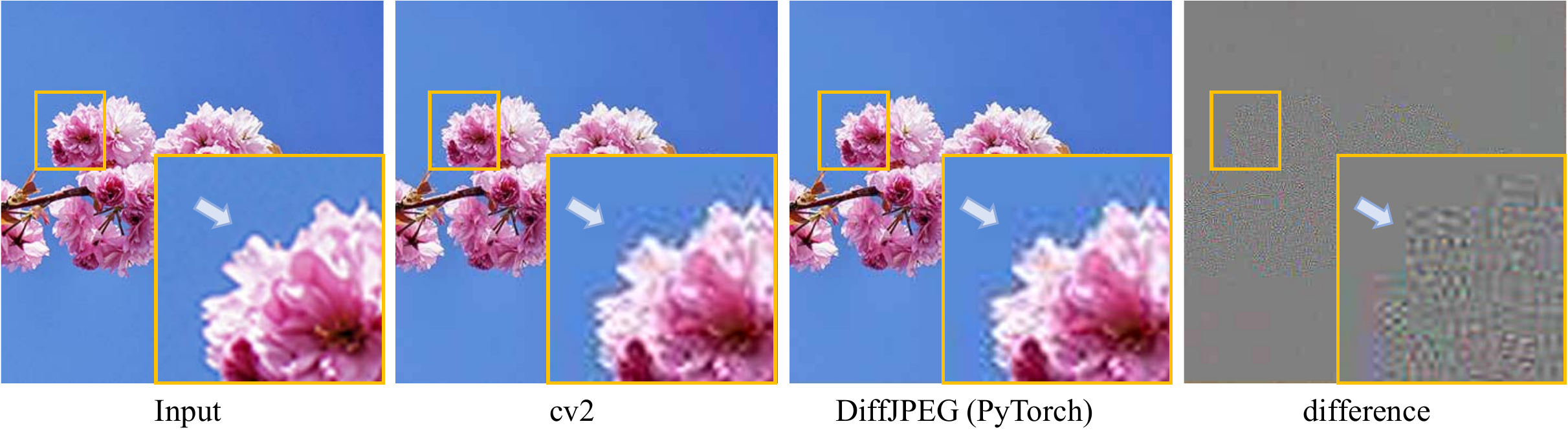}
	\end{center}
	\vspace{-0.5cm}
	\caption{JPEG compressed images by $\mathtt{cv2}$ and $\mathtt{DiffJPEG}$, with quality factor $q=50$. They produces slightly different results. Zoom in for best view}
	\label{fig:jpeg}
\end{figure}

We use the PyTorch implementation - $\mathtt{DiffJPEG}$.
We observe that the compressed images by $\mathtt{DiffJPEG}$ are a bit different from those compressed by the $\mathtt{cv2}$ package. Fig.~\ref{fig:jpeg} shows the typical JPEG compression artifacts and the difference caused by using different packages.
Such a difference may bring an extra gap between synthetic and real samples.
In this work,  we only adopt $\mathtt{DiffJPEG}$ for simplicity, and this difference will be addressed later.

\section{Quantitative Comparisons}\label{sec:quantitative}

We provide the non-reference image quality assessment - NIQE~\cite{mittal2013making} for reference. Note that existing metrics for perceptual quality cannot well reflect the actual human perceptual preferences on the fine-grained scale~\cite{blau20182018}.

We compare our Real-ESRGAN with several state-of-the-art methods, including ESRGAN~\cite{wang2018esrgan}, DAN~\cite{luo2020unfolding}, CDC~\cite{wei2020cdc}, RealSR~\cite{ji2020real} and BSRGAN~\cite{zhang2021designing}.
We test on several diverse testing datasets with real-world images, including RealSR~\cite{cai2019toward}, DRealSR~\cite{wei2020cdc}, OST300~\cite{wang2018sftgan}, DPED~\cite{ignatov2017dslr}, ImageNet validation~\cite{deng2009imagenet} and ADE20K validation~\cite{zhou2019semantic}.
The results are shown in Tab.~\ref{tab:qualitative}. Though our Real-ESRGAN+ does not optimize for NIQE scores, it sill produces lower NIQE scores on most testing datasets.
\section{More Qualitative Comparisons}\label{sec:more_results}
\begin{figure*}
	\vspace{-0.2cm}
	\begin{center}
		\includegraphics[width=1\linewidth]{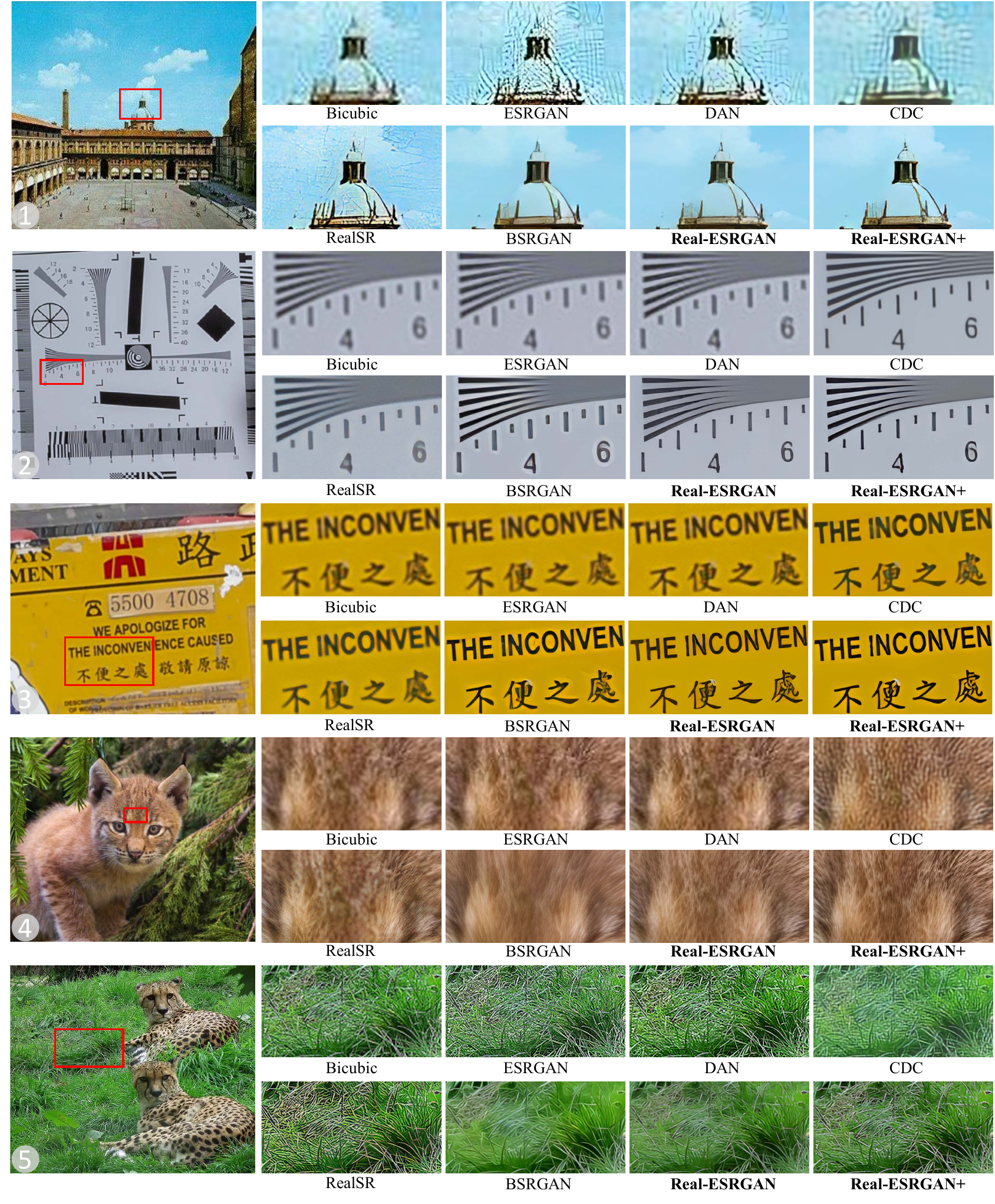}
	\end{center}
	\vspace{-0.5cm}
	\caption{Qualitative comparisons on several representative real-world samples with upsampling scale factor of 4. Our \mbox{Real-ESRGAN} outperforms previous approaches in both removing artifacts and restoring texture details. Real-ESRGAN+ (trained with sharpened ground-truths) can further boost visual sharpness.
		Other methods typically fail to remove complicated artifacts (the 1st sample) and overshoot artifacts (the 2nd, 3rd sample), or fail to restore realistic and natural textures for various scenes (the 4th, 5th samples).
		(\textbf{Zoom in for best view})}
	\label{fig:compare2}
\end{figure*}

We show more qualitative comparisons with previous works. As shown in Fig.~\ref{fig:compare2}, our \mbox{Real-ESRGAN} outperforms previous approaches in both removing artifacts and restoring texture details. Real-ESRGAN+ (trained with sharpened ground-truths) can further boost visual sharpness.
Other methods typically fail to remove complicated artifacts (the 1st sample) and overshoot artifacts (the 2nd, 3rd sample), or fail to restore realistic and natural textures for various scenes (the 4th, 5th samples).

